\documentstyle[12pt,epsf,subfigure]{article}
\newcommand{\cd}{\makebox[0.08cm]{$\cdot$}}
\title{Explicitly covariant light-front dynamics\\ and some its applications}

{\author{V.A. Karmanov$^a$\thanks{e-mail: karmanov@sci.lebedev.ru, 
karmanov@isn.in2p3.fr} \quad and $\;$ J. Carbonell$^b$\thanks{e-mail:
carbonel@isn.in2p3.fr} \\
{\small \em  $^a$Lebedev Physical Institute, Leninsky Prospekt 53, 119991
Moscow, Russia}\\
{\small \em $^b$Institut des Sciences Nucl\'eaires, 53 avenue des Martyrs,
38026 Grenoble, France}
}}

\begin{document}
\maketitle
\bibliographystyle{unsrt}

\begin{abstract}{
The light-front dynamics is an efficient approach to study of field
theory and of relativistic composite systems (nuclei at relativistic relative
nucleon momenta, hadrons in the quark models). The explicitly covariant version
of this approach is briefly reviewed and illustrated by some applications.}
\end{abstract}

\section{Introduction}\label{forms}

The recent experimental data on deuteron structure  and 
heavier nuclei concern the relativistic domain of the nucleon momenta. Their
successful theoretical description  was achieved \cite{ck99,antonov02} in the
framework of light-front dynamics (LFD). This approach deals with the wave
function defined on the four-dimensional space-time plane given by 
equation $t+z=0$. 

In general, the relativistic state vector is defined on a 
space-like surface. The  plane $t={\rm const}$, associated  with the
non-relativistic quantum mechanics, is a particular case of this surface. The
Schr\"odinger equation describes the dynamical time-evolution of the wave function
$\psi(\vec{r},t)$, or, in other words, from one space-like plane
$t={\rm const}$ to  other one. This dynamics was called by Dirac the "instant
form" \cite{dirac}. Light-front dynamics, also called "front form", is
associated with the plane $t+z=0$. The latter is considered as a limiting case
of a space-like plane. For completeness we mention the third form proposed by
Dirac: the "point form", associated with the hyperboloid
$x^2=t^2-\vec{r}^{\;2}={\rm const}$.

This paper is a brief review of the explicitly covariant version of light-front
dynamics. We explain its relation to the Bethe-Salpeter approach and illustrate
it by few examples. Namely, we give the wave function of two relativistic
scalar particles interacting by scalar exchange,  discuss  the relativistic
deuteron wave function and its electromagnetic form factors in comparison to
CEBAF/TJNAF data and present our new results concerning the system of three
relativistic bosons with zero-range interaction.

\subsection{Why light-front dynamics?}\label{why}

The theoretically self consistent description  of the relativistic composite
system is based on the quantum field theory. The state vector of a composite
system and
even of the vacuum has  a very complicated structure. It is usually described as
a superposition of bare quanta, corresponding to  non-interacting fields. If
we "switch off" the interaction between  fields, the number of particles is
conserved. As soon as we take into account the interaction, the state vector, is
in general a superposition of states with different numbers of particles.

When the interaction is weak, like in the case of quantum electrodynamics, it 
does not change the state vector too much
and the "dressed" electron differs from the bare one mainly  
by small admixture of photon.

The situation is drastically different, when the interaction is strong. For
example, the proton is usually treated as being a three quark system, 
but these quarks are not the same
than those appearing in the initial Lagrangian of Quantum Chromodynamics. They are
constituent quarks, which, in their turn, consist of bare quarks and
gluons. The proton state vector is a huge superposition of bare
fields and has not yet been calculated from the QCD first principles.

One should emphasize that not only the proton state, but also the state without
physical particles -- the vacuum state -- 
 is a complicated superposition of the bare particles, or, in other
words, of vacuum fluctuations. This description of
emptiness in terms of the very complicated conglomerates of particles seems
unnatural. It would be much better to use an approach, in  which the
vacuum is indeed nothing but emptiness. Simplifying the vacuum wave function,
we simplify the wave function of the dressed states as well,
eliminating from them, like in the vacuum wave function, the field fluctuations. 
After doing that one can study the real, physical structure of particles and
nuclei in a more higienic way.

{\em In the light-front dynamics the vacuum is nothing but emptiness}
and here is the main advantage of this approach.

Qualitatively it can be understood, by noticing that the front form
corresponds to the instant form, studied from a reference system
moving with velocity close to the speed of light. Indeed,
an observer moving with a
velocity $v$ along $z$-axis describes a physical process in his  coordinates
$(t',x',y',z')$, which are related to the laboratory ones by the Lorentz
transformation:  
$$
z'=\frac{z-vt}{\sqrt{1-v^2/c^2}},\quad 
t'=\frac{t+zv/c^2}{\sqrt{1-v^2/c^2}},\quad 
x'=x,\quad y'=y. $$

The plane $t'=0$ in the moving system corresponds to $t+zv/c^2=0$ in the
laboratory coordinates. 
In the limiting case, when $v\rightarrow c$, we get the light-front plane
$t'\propto z_+=t+z/c=0$. So, the front form  corresponds to the normal form 
$t'=0$, but studied from the system  of reference moving with the limiting
speed $c$ called since Weinberg \cite{weinberg} the "infinite momentum frame". 

Consider now the fluctuation creating three particles from vacuum (e.g., in the
$\varphi^3$-theory).  The fluctuation with energy  $\Delta E
=\varepsilon_{\vec{k}_1}+\varepsilon_{\vec{k}_2}+  \varepsilon_{\vec{k}_3}$ may
occur during the time $\Delta t\approx \hbar/\Delta E$ (here 
$\varepsilon_{\vec{k}}=\sqrt{\vec{k}\,^2+m^2}$). In the  infinite momentum
frame the momenta $\vec{k}_i$ and energies $\varepsilon_{\vec{k}_i}$ of any
particle increase and  $\Delta E$ tends to infinity. Therefore, the 
fluctuation time $\Delta t$ tends to zero. The contribution of this fluctuation  to
the vacuum wave function disappears.  Due to the time dilation, all 
physical processes are delayed, and  fluctuations have no time to occur. This
means that in a thought experiment in the infinite momentum frame we study
the particles prepared "far in advance", not spoiled by the vacuum
fluctuations.  From practical point of view, it is enough to work in a usual
reference frame, but with the state vector defined on the light-front plane. 

A disadvantage of this approach is the distinguished role of $z$-axis relative to
the $x,y$-axes. Because of that, the theory is not explicitly covariant.  To
overcome it, in the paper \cite{karm76} (see for review \cite{karm88,cdkm}) the
explicitly covariant version of light-front dynamics (CLFD) was proposed, which
deals with the state vector defined on the light-front plane $\omega\cd x=0$,
where $\omega= (\omega_0,\vec{\omega})$,
$\omega^2=\omega_0^2-\vec{\omega}^2=0$.  The theory becomes explicitly
covariant, whereas the dependence of the wave function on the light-front plane
is parametrized by means of the four-vector $\omega$. This is very convenient
from the point of view of applications. In particular the standard approach
is recovered  when $\omega=(1,0,0,-1)$.

\section{Wave function}\label{wfp} 
The wave functions studied in LFD
are the Fock components of the state
vector, that is, 
the coefficients in an expansion of the state vector $|p>$ with respect to the
basis of free fields:   
\begin{eqnarray}\label{wfp1} 
|p\rangle&\equiv& 
(2\pi)^{3/2}\int \psi_2(k_1,k_2,p,\omega\tau)
a^\dagger(\vec{k}_1)a^\dagger(\vec{k}_2)|0\rangle  \nonumber \\ &\times&
\delta^{(4)}(k_1+k_2-p-\omega\tau) 2(\omega\cd p)d\tau
\frac{d^3k_1}{(2\pi)^{3/2}\sqrt{2\varepsilon_{k_1}}}
\frac{d^3k_2}{(2\pi)^{3/2}\sqrt{2\varepsilon_{k_2}}}  + \cdots\ .  
\end{eqnarray} 
The dots include the higher Fock states. For simplicity, we omit the
spin indices. By construction, all the four-momenta are on the corresponding
mass shells: $k_i^2=m_i^2$, $p^2=M^2$, $(\omega\tau)^2=0$.

We emphasize the presence in (\ref{wfp1}) of the delta-function. 
It provides the conservation law:
\begin{equation}\label{sc1}                                                     
k_1+k_2=p+\omega\tau,                                                  
\end{equation} 
where $\tau$ is  a scalar variable responsible for the
off-energy shell continuation. The 
four-momentum $\omega\tau$ is associated, for convenience, with a fictive
particle called "spurion". 
We would like to emphasize at this point that
the use of "spurion" does not imply the use
of any unphysical degres of freedom in the theory
but only a confortable way of parametirsing the off-shell effects.
As mentioned above, the standard LFD  
corresponds to $\omega=(1,0,0,-1)$, or, in the light-front coordinates,  
$\omega_+=\omega_0+\omega_z=0$, $\omega_{\perp}\equiv\omega_{x,y}=0$,
$\omega_-=\omega_0-\omega_z=2$. In this case, 
equation (\ref{sc1})  gives
the standard conservation  laws for the $(\perp,+)$-components of the momenta
(which does not include any $\tau$),
but does not constrain the minus-components. Their difference
$k_{1-}+k_{2-}-p_-=2\tau$ determines the value of $\tau$.
Because of non-conservation of the minus-components, the
wave function is "off-energy shell". 

The term "off-energy shell" is borrowed from the old fashioned perturbation
theory, where it means that for  an amplitude which is an internal part of a
bigger diagram, there is no conservation law for the energies of the incoming
and outgoing particles. Note that any bound system is always off-energy shell,
since the sum of the constitutent energies $2\varepsilon_k$ (in the
c.m.-frame of two-body system) is not equal to  mass $M$ of the bound 
system but larger: $2\varepsilon_k\geq 2m> M$.
We emphasize the difference between the off-energy shell and  off-mass
shell amplitudes: in the latters the four-momenta squared are not equal to the
masses  squared, but satisfy the conservation law for all the four components.
An example of off-mass shell amplitude is the Bethe-Salpeter function (see 
below sect. \ref{bsf}).

As mentioned, the wave function, defined on the light-front plane, depends on
orientation of this plane. From (\ref{wfp1}) we see that this dependence is
given by the four-vector $\omega$. It is an important property of any Fock
component. At the same time, the observables, like form factors, don't depend
on $\omega$. In principle, the form facors, being scalars, could depend on the
scalar products  $F=F(q^2,\omega\cd p, \omega\cd p')$, where $q^2=(p-p')^2$.
Since $\omega$ determines a direction only, the theory is invariant relative to
the replacement $\omega\to a\omega$, where $a$ is a number. Therefore, the
extra scalar products should enter in the form of ratio: $\omega\cd p/\omega\cd
p'$. To factorize the e.m.-amplitude, one should impose the condition 
$\omega\cd q=0$ (equivalent to $q_+=0$ in the standard approach, see
\cite{cdkm} for details). With this condition we get $\omega\cd p/\omega\cd
p'=1$, so the form factor does  not depend on $\omega$. 

For the states with spins 1/2 and 1, decomposing the e.m. vertex in the form
factors, one can construct not only scalar products and not only usual spin
structures (two for spin 1/2 and three for spin 1), but also $\omega$-dependent
spin structures. The scalar coefficients at the front of them are extra,
non-physical form factors. One should separate the physical form factors from
non-physical ones. Corresponding formulas extracting the physical form factors
from full e.m. vertex were found in \cite{ks,km96}.

Parametrized in terms of the relative momentum $\vec{k}$, the two-body wave
function reads:
\begin{equation}\label{sc7}                                                     
\psi=\psi(\vec{k},\vec{n}),        
\end{equation}       
where $\vec{k}=\vec{k}_1$ in the system where $\vec{k}_1+\vec{k}_2=0$ and
$\vec{n}$ is the direction of $\vec{\omega}$ in this system.
We see that the relativistic light-front wave function
depends not only on the relative momentum $\vec{k}$ but on another variable --
the unit vector $\vec{n}$. 

It is also useful to intro\-duce another set of vari\-ables                                
\begin{equation}\label{sc8}
 x=\omega\cd k_1/\omega\cd p\ , \quad R=k_1-xp                                                                  
\end{equation}                                                                  
and represent the spatial part of the four-vector $R$ as 
$\vec{R}=\vec{R}_{\|} 
+\vec{R_{\perp}}$, where $\vec{R}_{\|}$ is parallel to $\vec{\omega}$  and 
$\vec{R}_{\perp}$ is orthogonal to $\vec{\omega}$.  Since 
$R\cd\omega=R_0\omega_0-\vec{R}_{\|}\cd\vec{\omega}=0$ by definition  of  $R$,
it follows that $R_0=|\vec{R}_{\|}|$, and, hence,  $\vec{R}^2_{\perp} =-R^2$ is
invariant. Therefore, $\vec{R}^2_{\perp}$  and $x$ can be chosen as
two scalar arguments of the wave function:                                                
\begin{equation}\label{sc9}                                                     
\psi=\psi(\vec{R}^2_{\perp},x) \ .                                         
\end{equation}                                                                  
Using the definitions of the variables $\vec{R}^2_{\perp}$ and $x$, we          
can readily relate them to $\vec{k}^2$ and $\vec{n}\cd\vec{k}$:                    
\begin{equation}\label{sc9a}                                                    
\vec{R}^2_{\perp}=\vec{k}\,^2-(\vec{n}\cd\vec{k})^2,\quad                          
x=\frac{1}{2}\left(1-\frac{\vec{n}\cd\vec{k}}{\varepsilon_k}\right).               
\end{equation}                                                                  
The variables introduced above can  be easily generalized to the case  of
different masses and an arbitrary number of  particles~\cite{karm88}. 

\section{Equation} \label{toto}                                       
                                                                                
\begin{figure}[!ht]
\centerline{\epsfbox{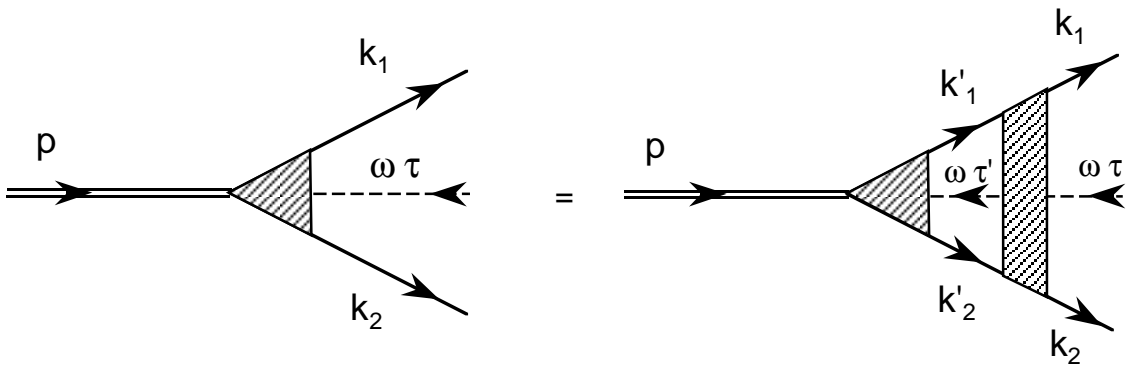}}
\caption{Equation for the two-body wave function.}
\label{feq}
\end{figure}

The equation for the wave function is shown graphically in  fig.~\ref{feq}. 
The dashed line corresponds to "spurion".  
In the case of a scalar particle, in 
terms of the variables $\vec{k}, \vec{n}$, this equation has the 
following form:  
\begin{equation}\label{eqwf3} 
\left(4(\vec{k}\,^2 + m^2)-M^2\right)\psi(\vec{k},\vec{n}) = 
-\frac{m^2}{2\pi^3} \int \psi(\vec{k}\,',\vec{n}) 
V(\vec{k}\,',\vec{k},\vec{n},M^2) \frac{d^3k'}{\varepsilon_{k'}} \ .  
\end{equation}                                                                  
The kernel of eq. (\ref{eqwf3}) depends on the vector variable $\vec{n}$ and on
the bound state mass $M^2$. This dependence is associated with the  retardation
of the interaction. 

In the  non-relativistic limit the light front wave function turns into the
ordinary equal-time wave function. This limit can be formally obtained at $c
\to \infty$. Then the light-front plane $t+z/c=0$ turns into the plane $t=0$.
Therefore the equation (\ref{eqwf3}) turns into the Schr\"odinger equation in
momentum space, the kernel $V$ being the Fourier-transform of non-relativistic
potential and the wave function no longer depends on $\vec{n}$.  

\section{Simple example}\label{simple}    
                                                   
For a given Lagrangian, the amplitudes are calculated by rules of special
graph technique, developed in \cite{kadysh} and adjusted to the light-front
dynamics  in \cite{karm76}. The diagrams correspond to the time-ordered graphs.
The amplitudes on energy-shell coincide with usual Feynman amplitudes on-mass
shell. Off the energy and mass shells, these amplitudes differ from each
other.

\begin{figure}[!ht]
\centerline{\epsfbox{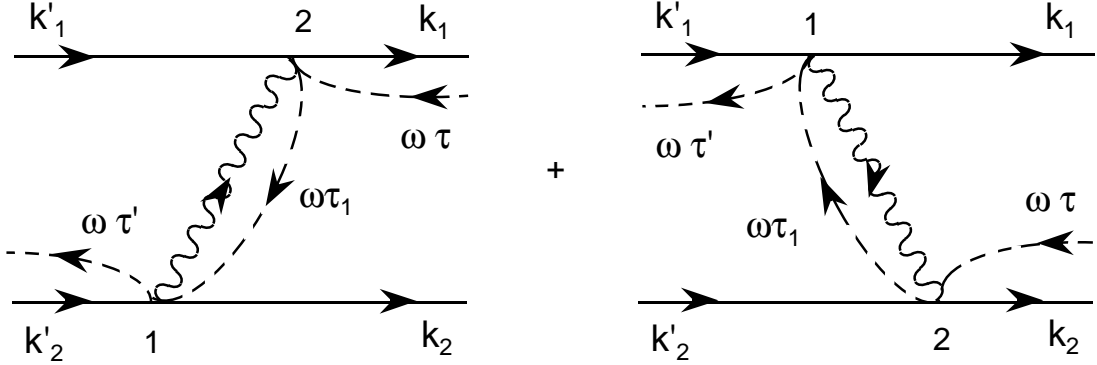}}
\caption{Exchange by a particle in $t$-channel.}
\label{fkern} 
\end{figure}

Consider exchange in the $t$ channel by a scalar particle of mass $\mu$ between
two scalar particles. The amplitude is given by two time-ordered diagrams shown
in fig.~\ref{fkern}. They determine, in the ladder approximation, the kernel of
the equation (\ref{eqwf3}). The external spurion lines indicate that the
amplitude is off-energy shell. 

According to the light-front graph technique, the amplitude has the form:   
\begin{equation}\label{k1} 
{\cal K}=
\frac{g^2\theta\left(\omega\cd 
(k_1-k'_1)\right)}{\mu^2-(k_1-k'_1)^2+2\tau 
\omega\cd(k_1-k'_1) -i\epsilon}
+\frac{g^2\theta\left(\omega\cd (k'_1-k_1)\right)}{\mu^2 
-(k'_1-k_1)^2 +2\tau' 
\omega\cd (k'_1-k_1)-i\epsilon}.  
\end{equation}
The two items in (\ref{k1}) correspond to the two diagrams of 
fig.~\ref{fkern}. 
They cannot be non-zero simultaneously. 
The off-energy shell terms, depending on $\omega$, appear automatically 
and they are explicitly given in (\ref{k1}).
On the energy shell,
i.e. for both $\tau=\tau '=0$,  the expression for the kernel becomes 
identical to the Feynman amplitude:                
\begin{equation}\label{k2}                                                      
{\cal K}(\tau=\tau'=0) = \frac{g^2}{\mu^2-(k_1-k'_1)^2-i\epsilon}\ .  
\end{equation} 

In the case of $\mu=0$ we obtain the Wick-Cutkosky model \cite{wcm}, 
but treated in the framework of CLFD \cite{karm80}. Going over from
the kernel ${\cal K}$ to  $V=-{\cal  K}/(4m^2)$, introducing the constant
$\alpha=g^2/(16\pi m^2)$, and  expressing (\ref{k1}) by means of the initial
and final relative momenta $\vec{k},\vec{k}'$, we obtain~\cite{karm80}:    
\begin{equation}\label{k3}                                                      
V=-\frac{4\pi\alpha}
{\displaystyle{(\vec{k}\,' - \vec{k}\,)^2 -
(\vec{n}\cd\vec{k}\,')(\vec{n}\cd\vec{k})                                           
\frac{(\varepsilon_{k'}-\varepsilon_k)^2}{\varepsilon_{k'}
\varepsilon_k} +(\varepsilon_{k'}^2
+\varepsilon_k^2-\frac{1}{2}M^2)                           
\left|\frac{\vec{n}\cd\vec{k}\,'}{\varepsilon_{k'}}                                
-\frac{\vec{n}\cd\vec{k}}{\varepsilon_k}\right|}}.
\end{equation}
The dependence on the light-front orientation manifests itself in (\ref{k3})
through the vector parameter $\vec{n}$. 

The solution of the equation (\ref{eqwf3}) with the kernel (\ref{k3}), 
in the limit of small binding energy, has the form:
\begin{equation}\label{k8}                                                     
\psi(\vec{k},\vec{n})                                                           
=\frac{8\sqrt{\pi m}\kappa^{5/2}}{(\vec{k}\,^2+\kappa^2)^2 
\left(\displaystyle{1+\frac{|\vec{n}\cd\vec{k}|}                                   
{\varepsilon_k}}\right)}\ .                                         
\end{equation}                                                                  
In non-relativistic limit $k,k'\ll m$ the kernel (\ref{k3})  turn into the
Coulombien one:  $V\approx -4\pi\alpha/(\vec{k}\,' - \vec{k}\,)^2$ (that in the
coordinate space corresponds to $V(r)=-\alpha/r$). The $\vec{n}$-dependent
factor in the wave function (\ref{k8}) disappears and the latter turns into the
Coulombien wave function too.

We emphasize that the dependence of the wave function (\ref{k8}) on $\vec{n}$
does not mean any violation of the rotational invariance, since this function
is a scalar, as should be for the zero angular momentum. However, for non-zero
angular momentum, because of Fock space truncation, one can find dependence of
energy on the projection of the angular momentum on the $\vec{n}$-direction.
This fact indeed means a violation of rotational invariance. 
A solution of this problem is given in \cite{mck2}.

The solutions of equation (\ref{eqwf3}) with the kernel (\ref{k3})  for $0\leq
\mu\leq m$ were studied in \cite{MC_PLB_00}. It was found that for  the
exchange mass $\mu$ comparable with the constituent mass $m$  the relativistic
effects in the binding energy are very important even for vanishingly small
("nonrelativistic") binding energy.

         
\section{Relation with the Bethe-Salpeter function}\label{bsf}                       

It is instructive to compare the solution (\ref{k8}) with one found
using the Bethe-Salpeter function \cite{bs}. 
The latter is defined as (see for review \cite{nak69}): 
\begin{equation} \label{bs2} 
\Phi(x_1,x_2,p)= \langle 0 \left|
T(\varphi (x_1)\varphi (x_2))\right|p\rangle .                                                                     
\end{equation}    
Here $\varphi (x)$ denotes the field operator in Heisenberg representation. The function (\ref{bs2})
depends on two four-vectors $x_{1},x_{2}$, they include two times 
$t_{1},t_{2}$. 
Though the Bethe-Salpeter function satisfies the normalization condition,
allowing to find the normalization coefficient, it  has no any probabilistic
interpretation.

In the
momentum space the Bethe-Salpeter function looks as: $\Phi=\Phi(q_1,q_2,p)$ 
and is defined through the Fourier transform:
\begin{eqnarray}\label{bs0}
\Phi(x_1,x_2,p)&=&(2\pi )^{-3/2}\exp \left[-ip\cd 
(x_1+x_2)/2\right]\tilde{\Phi}(x,p)\ ,\quad x=x_1-x_2\ ,                                              
\nonumber\\
\Phi(q_1,q_2,p)&=&\Phi (q,p)=\int \tilde \Phi (x,p)\exp (iq\cd x)d^4x\ ,                   
\end{eqnarray}                                                              
where $q=(q_1-q_2)/2$, $p=q_1+q_2$, $q_1$ and $q_2$ are off-mass  shell 
four-vectors: $q_1^2\neq m^2$, $q_2^2\neq m^2$.

In LFD the Heisenberg operators coincide with the Schr\"odinger ones 
on the light-front plane (like in the ordinary
formulation of field  theory the Heisenberg and Schr\"odinger operators are
identical for  $t=0$). Therefore, projecting the Bethe-Salpeter function
on the light-front, we can relate it to the light-front wave function. 
In the momentum space this relation reads \cite{cdkm}:
\begin{equation} \label{bs8}                                                     
\psi(k_1,k_2,p,\omega \tau ) =
\frac{(\omega\cd k_1 )(\omega\cd k_2 )}
{\pi (\omega\cd  p)}\int_{-\infty }^{+\infty }\Phi 
(q_1=k_1-\omega\tau/2+\omega\beta,q_2 
=k_2-\omega\tau/2-\omega\beta,p)d\beta       
\end{equation}      
where $\Phi(q_1,q_2)$ is the Bethe-Salpeter function (\ref{bs0}).
The argument $p$ in (\ref{bs8}) is
related to the on-shell  momenta $k_1,k_2$ as $p=k_1+k_2-\omega\tau$, that
follows from the off-mass shell relation $p=q_1+q_2$.

The equation (\ref{bs8}) makes the link between the Bethe-Salpeter function
$\Phi$ and the light-front wave function  $\psi$.  It should be  noticed
however that the wave function (\ref{bs8}) is not necessarily an exact solution
of  eq. (\ref{eqwf3}), since, as a rule, different approximations are made for
the Bethe-Salpeter kernel and for the light-front one. In the ladder
approximation, for example, the second iteration of the Bethe-Salpeter
amplitude contains the box diagram, including the time-ordered diagram with
two exchanged particles in the intermediate state.  This contribution is absent
in the light-front ladder kernel and in its higer order iterations.

We can check now that in the Wick-Cutkosky model 
the relation (\ref{bs8}) reproduces the wave function (\ref{k8}).
  In this model, 
the exact expression for the Bethe-Salpeter function 
 has the form  \cite{wcm,nak69}:
 \begin{equation}\label{wcm02}
\Phi(q,p)=-ic\left[\left(m^2-\frac{1}{2}M^2-q^2\right)
\left(m^2-(\frac{1}{2}p+q)^2-i0\right)
\left(m^2-(\frac{1}{2}p-q)^2-i0\right)\right]^{-1},
\end{equation}
where $c=2^5\sqrt{\pi m\kappa^5}$ with 
$\kappa=\sqrt{m|\epsilon_b|}= m\alpha/2$. 
 Substituting (\ref{wcm02}) in
(\ref{bs8}), we indeed reproduce the expression (\ref{k8}) 
for the light-front wave function. 

The e.m. form factors calculated through the light-front wave function and
through the Bethe-Salpeter one were compared in \cite{ks}. They coincide with
each other with high accuracy. This means that when
both approaches take into account the same relativistic dynamics, they give 
close results. 

However, an important difference between them is in the following.
Bethe-Salpeter function (\ref{bs2}) is a formally defined object. Whereas, the
light-front wave function, being the coefficient in the decomposition
(\ref{wfp1}), according to general principles of quantum theory, is the
probability amplitude. It is the same object as  nonrelativistic  wave function
and it has  the same physical meaning. Comparing light-front wave function of a
relativistic system with its non-relativistic counterpart, we describe  the
relativistic effects in a language of physical phenomena. This important point
is an advantage of LFD. Smooth matching of non-relativistic and relativistic
wave functions is also very convenient in practical calculations.

\section{Deuteron wave function and form factors}\label{spin} 

\begin{figure}[!ht] 
\begin{center}
\mbox{\epsfxsize=9.cm\epsffile{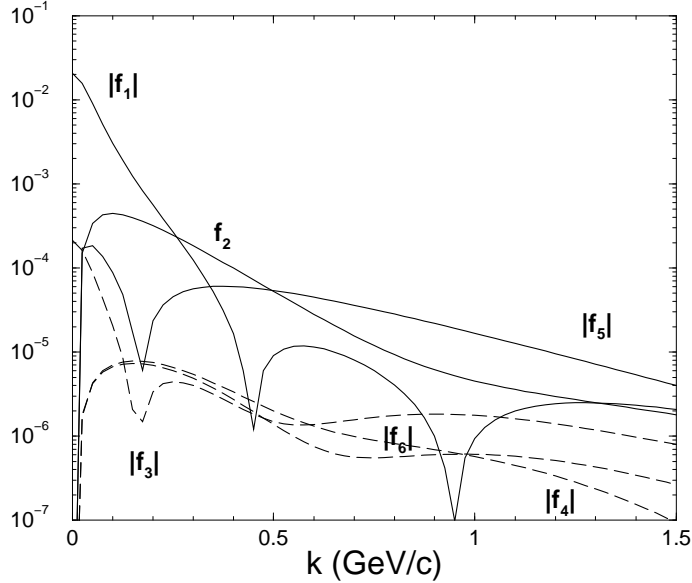}}
\end{center}
\caption{Deuteron relativistic components $f_1-f_6$ as a 
function of $k$, at $\vec{n} \vec{k}/k=0.6$.\label{fckd4}}
\end{figure}

The deuteron wave function
is determined by six invariant functions, depending on two
scalar variables. This number  is the dimension of the matrix depending on the
spin projections of the deuteron and two nucleons, divided by the factor 2 due
to the parity  conservation: $N=3\times 2 \times 2/2=6$.
This wave function reads:
\begin{eqnarray}\label{nz8} 
\vec{\psi}(\vec{k},\vec{n}) & = & f_1\frac{1}{\sqrt{2}}\vec{\sigma} 
+ f_2\frac{1}{2}\left(\frac{3\vec{k}(\vec{k}\cd\vec{\sigma})}{\vec{k}^2} 
-\vec{\sigma}\right) 
+ f_3\frac{1}{2}\left(3\vec{n}(\vec{n}\cd\vec{\sigma})
-\vec{\sigma}\right) 
 \\ 
& + & f_4\frac{1}{2k}\left(3\vec{k}(\vec{n}\cd\vec{\sigma}) 
+ 3\vec{n}(\vec{k}\cd\vec{\sigma}) - 2(\vec{k}\cd\vec{n})\vec{\sigma}\right) 
+  f_5\sqrt{\frac{3}{2}}\frac{i}{k}\left[\vec{k}\times \vec{n}\right] 
+ f_6\frac{\sqrt{3}}{2k}\left[[\vec{k}\times \vec{n}]\times\vec{\sigma}\right].\nonumber
\end{eqnarray} 
In the relativistic  one boson exchange model it was calculated in
\cite{ck-deut}.  The results are displayed in fig. \ref{fckd4}.
One can see that the function $f_5$, of relativistic origin, is very important:
it dominates at $k > 500$ MeV/c. In \cite{dkm95} it was shown that this
component  dominates since it  automatically takes into account the isovector
exchange currents (so called pair terms). They should not be added separately.
In nonrelativistic limit the functions $f_{3-6}$ become negligible and only
two first functions $f_{1,2}$, corresponding to usual S- and D-waves,
survive.

\begin{figure}[p]
\begin{minipage}{14.cm}
\vspace{-2.5cm}
\begin{center}
\mbox{\epsfxsize=7.5cm\epsffile{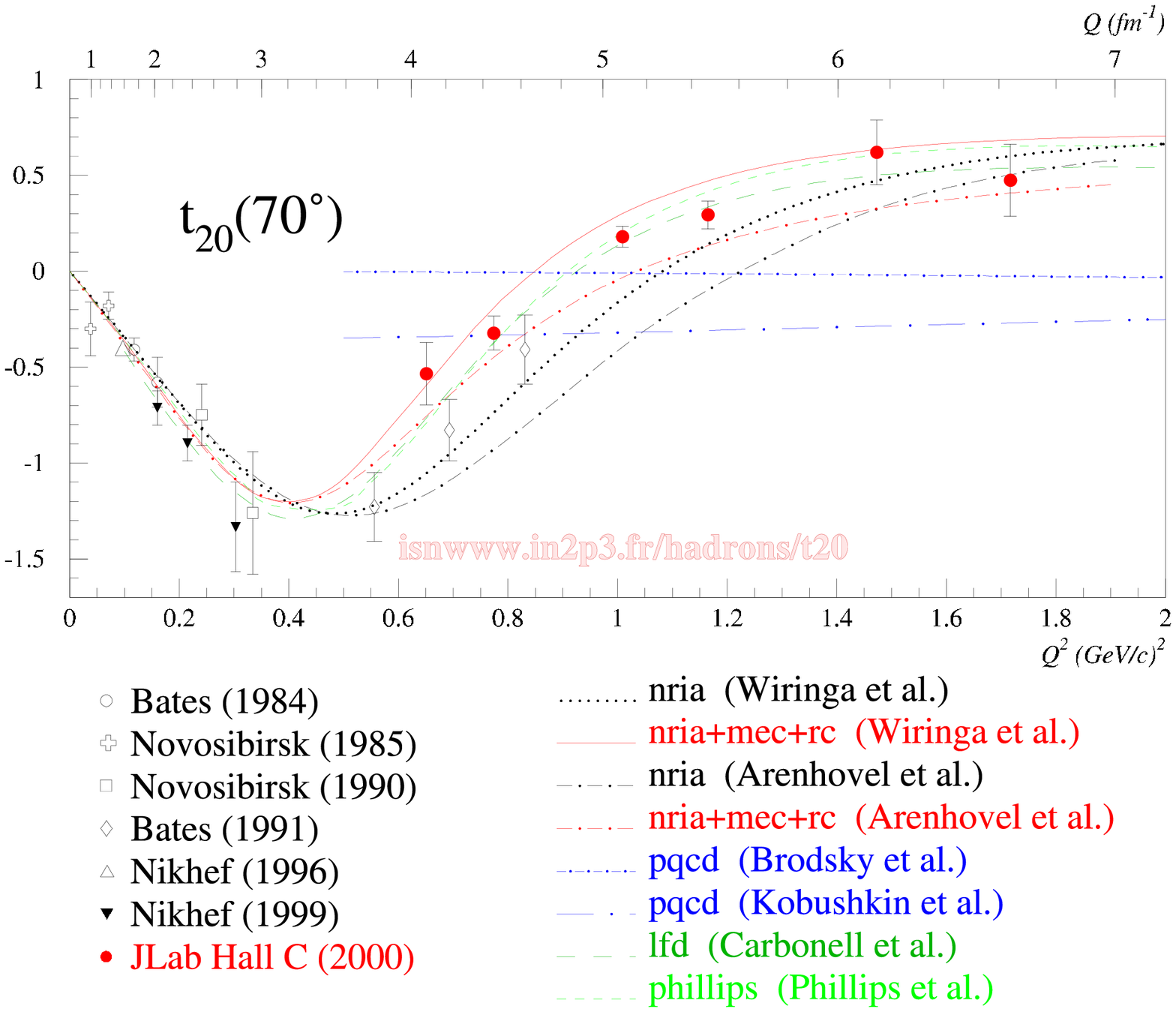}}
\vspace{-2.cm}
\caption{Polarization observable $t_{20}$ in $ed$ scattering, taken from 
\cite{wwwisn}.}\label{ft20}
\end{center}
\end{minipage}

\vspace{1.5cm}

\begin{minipage}{14.cm}
\begin{center}
\mbox{\epsfxsize=7.5cm\epsffile{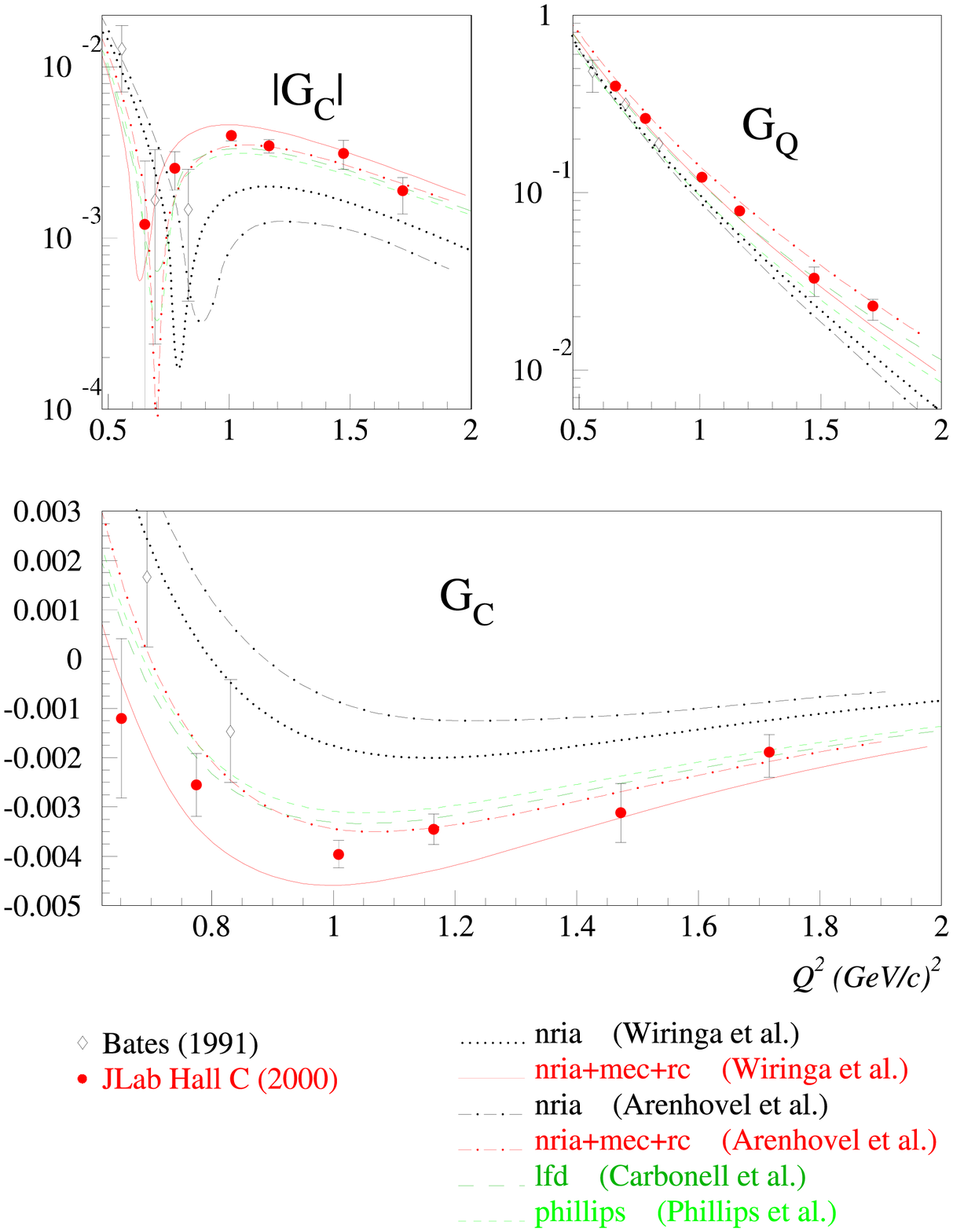}}
\vspace{0.5cm}
\caption{Charge and quadrupole deuteron form factors, taken from 
\cite{wwwisn}.\label{fgcgq}}
\end{center}
\end{minipage}
\end{figure}

This wave function was used in \cite{ck99,cdkm} to calculate the
deuteron electromagnetic form factors. No any parameters were fitted.  It
turned out that the calculated polarization observable $t_{20}$, fig.
\ref{ft20}, as well as the charge and   quadrupole form factors, fig.
\ref{fgcgq}, coincide with rather precise experimental data obtained recently
at CEBAF/TJNAF \cite{t20}.

\vspace{0.0cm}

\section{Three-boson bound states with zero-range
interaction}\label{3bosons}

\begin{figure}[!ht]
\begin{center}
\mbox{\epsfxsize=14.cm\epsffile{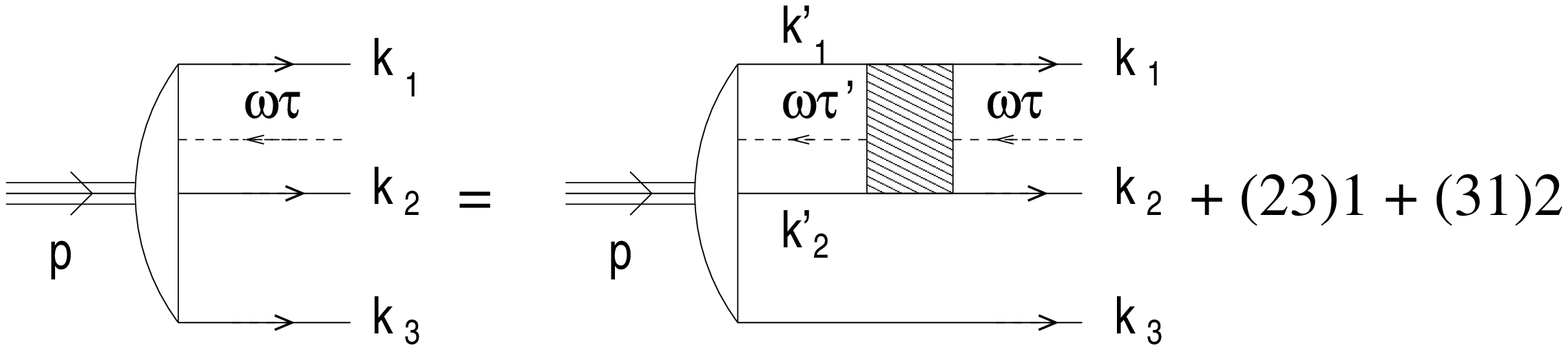}}
\caption{Three-body equation for the vertex function $\Gamma$.\label{fig1}}
\end{center}
\end{figure}

The most recent application of CLFD is the relativistic system of
three-bosons with zero range interaction. We will describe it in more detail.

Zero-range two-body interaction provides an important limiting case which
qualitatively reflects characteristic properties of nuclear and
atomic few-body systems. In the nonrelativistic three-body system
it generates the Thomas collapse \cite{thomas}. The latter means that the
three-body binding energy tends to $-\infty$, when the interaction radius tends
to zero. 

When the binding energy or the exchanged particle mass is not negligible in
comparison to the constituent masses, the nonrelativistic treatment becomes
invalid and must be replaced by a relativistic one. Two-body calculations show
that in the scalar case, relativistic effects are repulsive (see e.g.
\cite{MC_PLB_00}). Relativistic three-body calculations with zero-range
interaction have been performed in a  minimal relativistic model \cite{noyes}
and in the framework of the Light-Front Dynamics \cite{tobias}. In these works
it was concluded that, due to relativistic repulsion, the three-body binding
energy remains finite and the Thomas collapse is consequently avoided. However,
in the papers \cite{noyes,tobias} a cutoff has been introduced implicitly.
Being interested in studing the zero-range interaction as an important limiting
case, we eliminate this cutoff.

The three-body equation is represented graphically in figure \ref{fig1}.
It concerns the vertex function $\Gamma$,
related to the wave function $\psi$ in the standard way:
$$
\psi(k_1,k_2,k_3,p,\omega\tau)=\frac{\Gamma(k_1,k_2,k_3,p,\omega\tau)}
{{\cal M}^2-M^2_3},\quad
{\cal M}^2=(k_1+k_2+k_3)^2=(p+\omega\tau)^2.
$$
Like for the two-body system, 
all four-momenta are on the corresponding mass shells: $k_i^2=m^2,
p^2=M_3^2,(\omega\tau)^2=0$ and satisfy the conservation law
$k_1+k_2+k_3=p+\omega\tau$ involving  $\omega\tau$.

Applying to figure \ref{fig1} the covariant light-front graph techniques 
\cite{cdkm}, we find:
\begin{eqnarray}\label{eq3}
&&\Gamma(k_1,k_2,k_3,p,\omega\tau)=
\frac{\lambda}{(2\pi)^3} \int \Gamma(k'_1,k'_2,k_3,p,\omega\tau')
\nonumber\\
&&\times
\delta^{(4)}(k'_1+k'_2-\omega\tau'-k_1-k_2+\omega\tau)
\frac{d\tau'}{\tau'}\,
\frac{d^3k'_1}{2\varepsilon_{k'_1}}\,
\frac{d^3k'_2}{2\varepsilon_{k'_2}}
\;+\; (23)1+(31)2,
\end{eqnarray}
where  $\varepsilon_{k}=\sqrt{m^2+\vec{k}^2}$.
For zero-range forces, the interaction kernel is replaced 
in (\ref{eq3}) by a constant $\lambda$.
The contribution of interacting pair 12 is explicitly written while
the contributions of the remaining pairs are denoted by $(23)1+(31)2$.

The Faddeev amplitudes $\Gamma_{ij}$ are introduced in the standard way:
$$\Gamma(1,2,3)=\Gamma_{12}(1,2,3)+\Gamma_{23}(1,2,3)+\Gamma_{31}(1,2,3), $$
and equation (\ref{eq3}) is equivalent to a system of three coupled equations.
With the symmetry relations $\Gamma_{23}(1,2,3)=\Gamma_{12}(2,3,1)$ and
$\Gamma_{31}(1,2,3)=\Gamma_{12}(3,1,2)$, the system is reduced to a single
equation for one of the amplitudes,  say $\Gamma_{12}$.

We use the tree-body version of the variables defined in eq. (\ref{sc8}).  In
general, $\Gamma_{12}$ depends on all  variables ($\vec{R}_{i\perp},x_i$),
constrained by the relations
$\vec{R}_{1\perp}+\vec{R}_{2\perp}+\vec{R}_{3\perp}=0$, $x_1+x_2+x_3=1$, but
for  contact kernel it depends only on $(\vec{R}_{3\perp},x_3)$ \cite{tobias}.

The value of $\lambda$ is found analytically 
by solving the two-body problem with the same
zero-range interaction under the condition that the two-body bound state mass
has a fixed value $M_{2}$. In this way, the constant $\lambda$ disappears from
the kernel, but the two-body scattering amplitudes appears, as in any Faddeev
equations. For the zero-range interaction, this amplitude does not depend on 
the relative momentum, but depends on the off-energy shell variables. 
Therefore it is extracted from the integral.
Finally, the equation for the Faddeev amplitude reads:
\begin{equation}\label{eq29a}
\Gamma_{12}(R_{\perp},x)
=F(M_{12})\frac{\displaystyle{1}}{\displaystyle{(2\pi)^3}}
\displaystyle{\int_0^1}
\displaystyle{dx'}
\displaystyle{\int_0^{\infty}}
\frac{\Gamma_{12}\left(R'_{\perp},x'(1-x)\right)\;d^2R'_{\perp}}
{\displaystyle{(\vec{R'}_{\perp}-x'\vec{R}_{\perp})^2+m^2-x'(1-x')M_{12}^2}},
\end{equation}
where $F(M_{12})$ is the two-body off-shell scattering amplitude:
$$
\frac{F(M_{12})}{8\pi^2}=
\left\{
\begin{array}{l}
\left[\frac{\displaystyle{1}}
{\displaystyle{2y'_{M_{12}}}}\log \frac{\displaystyle{1+y'_{M_{12}}}}
{\displaystyle{1-y'_{M_{12}}}}
-\frac{\displaystyle{\arctan y_{M_{2}}}}
{\displaystyle{y_{M_{2}}}}\right]^{-1},\quad
M_{12}^2<0,\quad
y'_{M_{12}}=\frac{\sqrt{-M_{12}^2}}{\sqrt{4m^2-M_{12}^2}}
\\
\\
\left[\frac{\displaystyle{\arctan y_{M_{12}}}}
{\displaystyle{y_{M_{12}}}}
-\frac{\displaystyle{\arctan y_{M_{2}}}}{\displaystyle{y_{M_{2}}}}\right]^{-1},
\quad 0\leq M_{12}^2<4m^2,\quad 
y_{M_{12}}=\frac{M_{12}}{\sqrt{4m^2-M_{12}^2}}
\end{array}
\right.
$$
and similarly for $y_{M_{2}}=\frac{M_{2}}{\sqrt{4m^2-M_{2}^2}}$.
The variable
$M^2_{12}=(k'_1+k'_2-\omega\tau')^2=(p-k_3)^2$
is the off-shell mass of the two-body subsystem.
It is expressed through $M_3^2,R_{\perp}^2,x$ as
\begin{equation}\label{eq17}
M^2_{12}=(1-x)M_3^2-\frac{R_{\perp}^2+(1-x)m^2}{x}.
\end{equation}
The three-body mass $M_3$ enters in the  equation (\ref{eq29a}) 
through variable $M_{12}^2$.

By replacing  $x'(1-x)\to x'$, equation (\ref{eq29a}) can be transformed to 
\begin{equation}\label{eq30}
\Gamma_{12}(R_{\perp},x)
=F(M_{12})\frac{\displaystyle{1}}{\displaystyle{(2\pi)^3}}
\displaystyle{\int_0^{1-x}}
\frac{\displaystyle{dx'}}{\displaystyle{x'(1-x-x')}}\;
\displaystyle{\int_0^{\infty}}
\frac{\displaystyle{d^2R'_{\perp}}}{\displaystyle{{{\cal M}'}^2-M_3^2}}\;
\Gamma_{12}\left(R'_{\perp},x'\right),
\end{equation}
where
\[ {{\cal M}'}^2=\frac{\vec{R'}^2_{\perp}+m^2}{x'}
+\frac{\vec{R}^2_{\perp}+m^2}{x}
+\frac{(\vec{R'}_{\perp}+\vec{R}_{\perp})^2+m^2}{1-x-x'}. \]

This equation is the same than equation (11) from \cite{tobias} except
for the integration limits of ($\vec{R'}_{\perp},x'$) variables.
In \cite{tobias} the integration limits follow from the condition $M_{12}^2>0$.
They read
\begin{equation}\label{FBC}
\int_{m^2\over M_3^2}^{1-x} \left[\ldots\right] dx'
\int_0^{k^{max}_{\perp}} \left[\ldots\right]  d^2R'_{\perp}
\end{equation}
with $k^{max}_{\perp}=\sqrt{(1-x')(M_3^2x'-m^2)}$
and introduce a lower bound on the three-body mass $M_3>\sqrt{2}m$.
The same condition,
though in a different relativistic approach, was used in \cite{noyes}.
The integration limits in (\ref{FBC}) restrict the arguments
of $\Gamma_{12}$ to the domain
\[ {m^2\over M_3^2} \leq x \leq 1 -{m^2\over M_3^2} ,
\quad 0\leq R_{\perp}\leq k_{\perp}^{max}\]
and can be considered as a  method of regularization.
In this case, one no longer deals with the zero-range forces.

\begin{figure}[!ht]
\begin{center}
\epsfxsize=8.8cm\subfigure[ ]{\epsffile{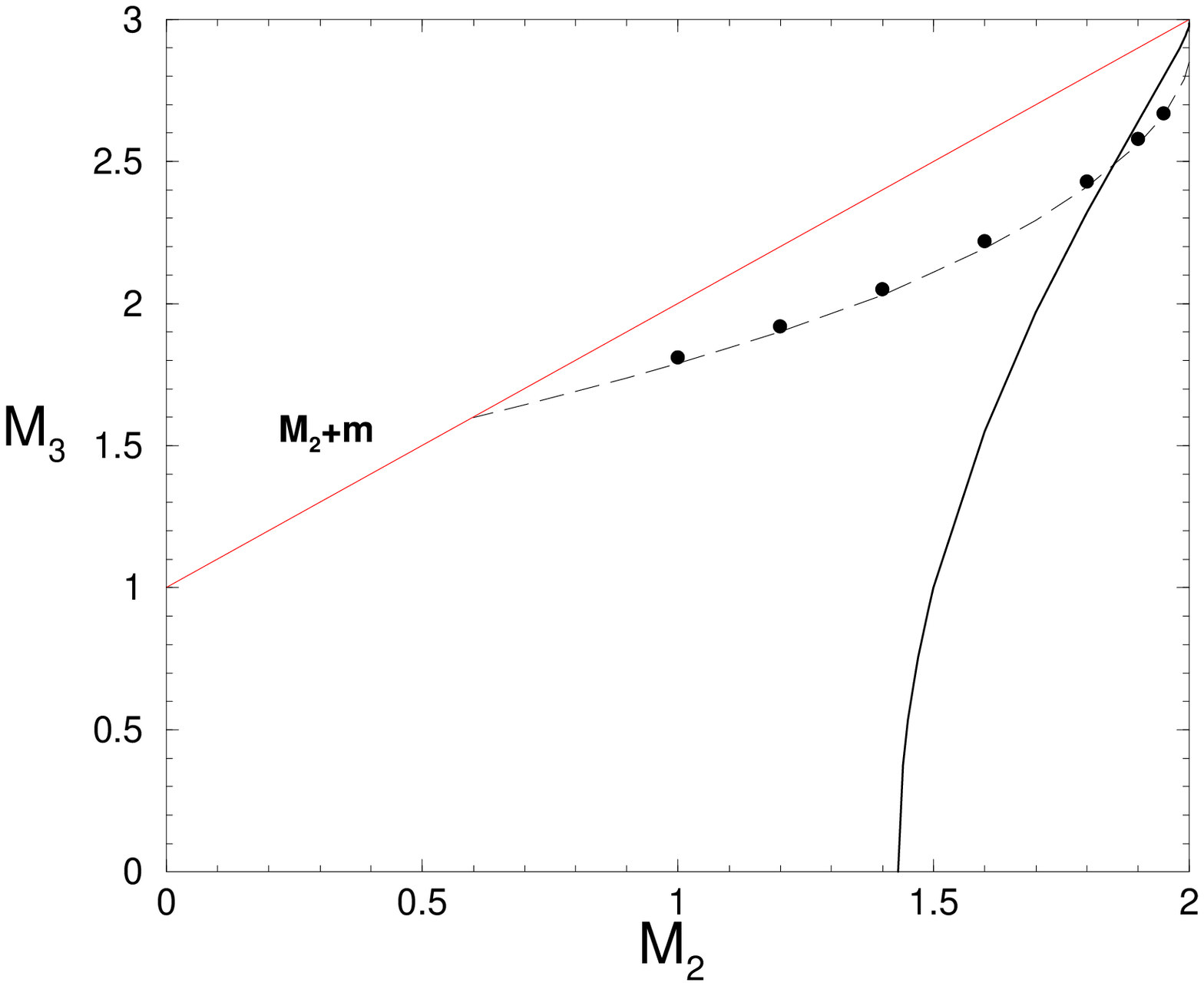}}\hspace{0.5cm}
\epsfxsize=6.8cm\subfigure[ ]{\epsffile{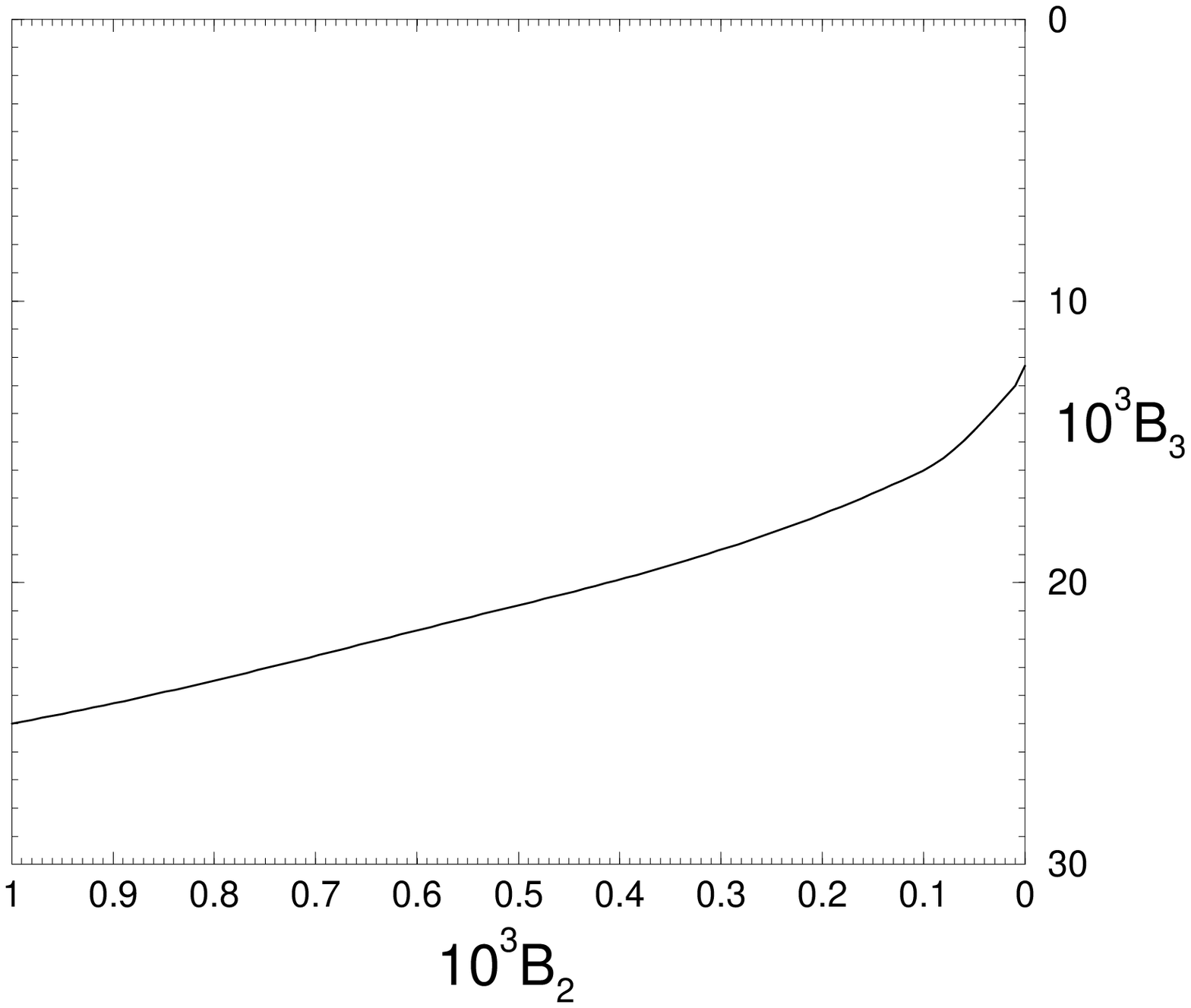}}
\caption{(a) Three-body bound state mass $M_3$ versus
the two-body one $M_{2}$ (solid line). 
Results  obtained with
integration limits (\protect{\ref{FBC}}) are in dash line.
Dots values are taken from \cite{AMF_PRC_95}.
(b) Zoom of the zero two-body binding region ($M_2\to2m,B_2\to0$)
displaying solid line only.}\label{Fig_M3_M2}
\end{center}
\end{figure}

The results of solving equation (\ref{eq29a}) are
presented in what follows. Calculations were carried out with constituent mass 
$m=1$ and correspond to the ground state.
We represent in fig. \ref{Fig_M3_M2}a
the three-body bound state mass $M_3$ as a function of the two-body one $M_2$
(solid line) together with the dissociation limit $M_3=M_2+m$.
The zero two-body binding limit $B_2=2m-M_2\to0$
is magnified in fig. \ref{Fig_M3_M2}b. In this limit
the three-boson system has a binding energy  $B_3\approx0.012$.

When $M_2$ decreases, the three-body mass $M_3$  decreases very
quickly and vanishes at the two-body mass value $M_{2}=M_{c}\approx 1.43$.
Whereas the meaning of collapse as used in the Thomas paper implies unbounded
nonrelativistic binding energies and cannot be used here,
the zero bound state mass $M_3=0$ constitutes its relativistic counterpart.
Indeed, for two-body
masses below the critical value $M_c$, the three-body system  no longer exists.
\begin{figure}[!ht]
\begin{center}
\mbox{\epsfxsize=7.cm\epsffile{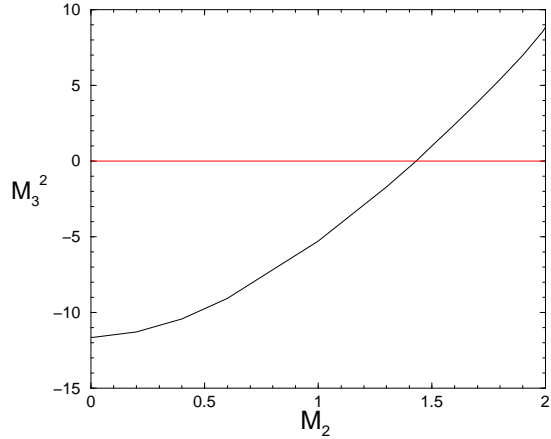}}
\caption{Three-body bound state mass squared $M_3^2$ versus $M_{2}$.}
\label{Fig_M32_M2}
\end{center}
\end{figure}

The results corresponding to integration limits (\ref{FBC}) are included in
fig. \ref{Fig_M3_M2}a (dash line) for comparison. Values given in \cite{tobias}
were not fully converged \cite{PC}. They have been corrected in
\cite{AMF_PRC_95} and are indicated by dots. In both cases the repulsive
relativistic effects produce a natural cutoff in equation (\ref{eq29a}),
leading to a finite spectrum and -- in the Thomas sense -- an absence of
collapse, like it was already found in \cite{noyes}. However, except in the
zero binding limit, solid and dash curves strongly differ from each other.

We would like to remark that for $M_2\leq M_c$, equation (\ref{eq29a}) posses
square integrable solutions with negative values of $M_3^2$. They have no
physical meaning but $M_3^2$ remains finite in all the two-body mass range
$M_2\in[0,2]$. The results of $M_3^2$ are given in figure \ref{Fig_M32_M2}.
When $M_{2}\to 0$,  $M_3^2$ tends to $\approx -11.6$.

From the above results we conclude
that the three-body bound state exists for two-body mass values
in the range $M_c\approx 1.43\,m \leq M_{2}\leq 2\,m$. At the zero two-body
binding limit, the three-body binding energy is $B_3\approx 0.012\,m$. The
Thomas collapse is avoided in the sense that three-body mass $M_3$  is finite,
in agreement with \cite{noyes,tobias}. However, another kind of catastrophe
happens. Removing infinite binding energies, the relativistic dynamics
generates zero three-body mass $M_3$ at a critical value  $M_2=M_c$. For
stronger interaction, i.e. when $0\leq M_{2}< M_c$,  there are no physical
solutions with real value of $M_3$. In this domain, $M_3^2$ becomes negative
and the three-boson system cannot be described by zero range forces, as it
happens in nonrelativistic dynamics. This fact can be interpreted as a
relativistic collapse.

\section{Conclusion}

We  have seen that the explicitly covariant light-front dynamics
provides a very satisfactory framework,
both from a theoretical and practical points of
view, for describing  relativistic nucleons in deuteron and other nuclei.  
The predicted deuteron
electromagnetic form factors, calculated  without any fitting
parameters, are in good agreement with experimental data.

This approach is also successfully when applied to many other problems
and a consequent effort covering 
several  directions of research is actually being performed.
Here is a list of current applications.
\begin{itemize}
\item
Deuteron wave function \cite{ck-deut} and  form factors \cite{ck99,cdkm}.
\item
Nucleon momentum distributions in nuclei \cite{antonov02}.
\item
Leptonic decay of heavy quarkonia \cite{bdm}.
\item
Fock-space expansion \cite{sbk98}.
\item
Nucleon wave function  \cite{karm98} and neutron charge radius \cite{ncr} in quark model.
\item
Two-fermion relativistic bound states \cite{yuk,cmk02}.
\item
Perturbative \cite{dkm01} and non perturbative \cite{bckm}  renormalization.
\item
Three-boson relativistic bound states \cite{ck3b}.
\end{itemize}
We conclude that the explicitly covariant light-front dynamics is a rather
efficient approach to relativistic few-body systems and to  field theory.

\bigskip
This work is supported by the French-Russian PICS and RFBR grants
Nos. 1172 and 01-02-22002.


\end{document}